\documentclass[aps,floats,floatfix,amssymb,amsmath,reprint,nofootinbib,prl,twocolumn,superscriptaddress]{revtex4-1}

\usepackage{dcolumn}
\usepackage{graphicx}
\usepackage{color}

\usepackage[normalem]{ulem}


\begin{document}
\title{Non-Fermi-liquid fixed point in multi-orbital Kondo impurity model relevant for Hund's metals}
\author{Alen Horvat}
\affiliation{Jozef Stefan Institute, Jamova 39, SI-1000, Ljubljana, Slovenia}
\author{Rok \v{Z}itko}
\affiliation{Jozef Stefan Institute, Jamova 39, SI-1000, Ljubljana, Slovenia}
\affiliation{University of Ljubljana, Faculty of Mathematics and Physics,  Jadranska 19, Ljubljana, Slovenia}
\author{Jernej Mravlje}
\affiliation{Jozef Stefan Institute, Jamova 39, SI-1000, Ljubljana, Slovenia}

\date{\today}
\begin{abstract}
 Due to the separation between the spin and the orbital screening
 scales, the normal state of Hund's metals at ambient temperature can
 be loosely characterized as a partially coherent state with
 fluctuating spins and quenched orbital moments. With the aim to
 characterize this situation  more precisely, we investigate the
 Kondo-Kanamori impurity model that describes the low-energy local
 physics of three-orbital Hund's metals occupied by two or four
 electrons. Within this model one can diminish the mixed spin-orbital
 terms and thereby enhance the separation between the two screening
 scales, allowing a more precise investigation of the intermediate
 state.
Using the numerical renormalization group we calculate the impurity
entropy as well as the temperature and frequency dependence of
the spin and the orbital susceptibilities. We  uncover a
non-Fermi-liquid  two-channel overscreened SU(3) fixed point that controls the behavior
in the intermediate regime. We discuss its fingerprints in the frequency
dependence of local orbital susceptibility and the shape of the
spectral function.
\end{abstract}
\maketitle

Ruthenates exhibit remarkable properties such as bad-metallic behavior
at high temperatures, a small value of temperature below which a
Fermi-liquid behavior is observed in measurements of
transport~\cite{tyler98}, and unusual optical response~\cite{lee02,
  capogna02,kamal06,schneider14,Georges2013}. Ruthenates have four
electrons in the $t_{2g}$ shell of extended 4d orbitals that
experience only moderate Coulomb repulsion, hence the occurrence of
correlations signified by a low-coherence scale and sizeable
quasiparticle renormalizations was considered
mysterious~\cite{capogna02}.  The dynamical mean-field theory
(DMFT)~\cite{georges_review_dmft} calculations that map the bulk
problem to a problem of a quantum impurity in an effective bath have
related the occurrence of the low coherence scale~\cite{haule09} and
related spin-freezing behavior~\cite{werner08} to strong electronic
correlations that are caused by the Hund's
coupling~\cite{mravlje11,Medici2011,Georges2013}. The same Hund's
physics applies also to iron
pnictides~\cite{haule09,Yin_kinetic_frustration_allFeSC}. A term
``Hund's metals'' has been introduced to describe ruthenates, iron
pnictides, as well as related compounds~\cite{isidori19}, and a term
Hund's impurity was proposed for multiorbital impurities on metallic
hosts~\cite{Khajetoorians2015,li_14}.

A successful line of thinking associates the correlations in
Hund's metals  with the proximity to a half-filled
Mott insulator~\cite{ishida10,misawa12,medici14,steinbauer19}:  the Hund's 
coupling favors large spin and blocks charge fluctuations of the half
filled ground state.  A different and equally successfully line of
thinking considers Hund's metals in terms of the low-energy
Kondo physics of the effective impurity model obtained by the DMFT mapping, which is a subject this
paper will elaborate on, too.   
 This point of view is based on the
observation that the Hund's coupling suppresses the spin-coherence
temperature~\cite{Yin2012,Aron2015,Stadler2015,Horvat2016,Mravlje2016,horvat2017,Stadler2019,deng19}. This scale
suppression is related to a reduced Kondo coupling constant for
the spin degree of freedom that occurs because the Hund's interaction
favors those charge fluctuations where the added electron is
parallel, which competes with the usual anti-ferromagnetic Kondo
coupling driven by the Pauli principle. The orbital-Kondo couplings
are,  meanwhile, not affected by the Hund's
coupling~\cite{Yin2012,Aron2015,Horvat2016}.  The Hund's coupling
hence leads to a distinct Kondo screening scale for spins and
orbitals, $T_K^{S}, T_K^{{L}}$, respectively, 
which was first suggested in Ref.~\cite{okada73}.

The intermediate-temperature state where $T_K^{S} < T < T_K^{L}$ has
been characterized in the
literature~\cite{Yin2012,Aron2015,Stadler2015,Horvat2016,Mravlje2016,horvat2017,Stadler2019,deng19}
as a state with fluctuating spins with a Curie dependence of spin
susceptibility, $\chi_{S} \propto 1/T$, and quenched orbitals with
constant orbital susceptibility, $\chi_{L} \sim \mathrm{const}$. 
The spin and the orbital screening scales are however not 
fully independent  because of the mixed spin-orbital coupling
terms $J_{ls}$ (see below for a precise definition).  The Hund's
metals are, in fact, characterized by a slow two-stage
crossover to a fully screened Fermi liquid. It is a key question
whether the intermediate state can be characterized in a more precise
way and what is the expected behavior of the observables there. Namely,
within the DMFT picture  the solid is a collection of atoms
having  some high-energy multiplets that are self-consistently
screened/quenched as we flow to low energy. The key issue here
is whether during that flow one passes close to some non-trivial
fixed point (and what that fixed point is) or whether one goes
directly into the Fermi liquid  ground state.

To address this, in this paper we consider a Kondo model relevant to
a  Hund's impurity. This allows us to
suppress $J_{ls}$ and separate the spin- and the orbital screening
scales far from each other.  In the intermediate temperature regime we
reveal a non-Fermi liquid behavior that can be associated to an
overscreened two-channel SU(3) fixed point. The orbital susceptibility
behaves with frequency as $\omega^{1/5}$ in this regime. We discuss
also the implications of this regime for the shape of the spectral
function and discuss the relevance of our findings for the physics of
Hund's metals.

The relevance of non-Fermi-liquid physics for the ruthenates within
the DMFT description was first discussed in Ref.~\cite{werner08}.  In
contrast to that paper that suggested the non-Fermi-liquid physics to
persist to zero temperature, which turned out not to be the case, we
stress that the NFL physics revealed here applies to the incoherent
regime, only. 

 Multi-orbital impurities with largely quenched orbital degrees of
freedom but fluctuating spins are equally relevant in the context of
magnetic adsorbates on surfaces. These can be probed at the
single-atom level using scanning tunneling microscopy and
spectroscopy, providing a direct way of probing local non-Fermi-liquid
phenomena through characteristic spectral features~\cite{Khajetoorians2015,Hiraoka2017}.


{\it Model and methods --}
We study the three-orbital impurity occupied  by two
electrons. The Anderson interaction term reads
\begin{equation}
  H_{\rm int}= (U-3J) \frac{\hat{N}(\hat{N}-1)}{2}  
   -2J \mathbf{S}^2 -\frac{1}{2}J \mathbf{L}^2.
\label{eq:hami}
\end{equation}
Here $U$ is the Coulomb repulsion, $J$ the Hund's coupling, $\hat{N}$
the charge operator, $\mathbf{S}$ the spin operator and $\mathbf{L}$
the orbital angular momentum operator. Eliminating the charge
fluctuations and taking into account that Hund's rule coupling
binds the two electrons at the impurity into a spin $S=1$ and orbital
momentum $L=1$ object, this model maps onto a Kondo Hamiltonian
\begin{eqnarray} 
	\label{Hkk0}
	H_{K} =  H_0+ J_{s}  \mathbf{S \cdot \sigma} + J_{l}  \mathbf{L \cdot l}
	+ J_{q}  \mathbf{Q \cdot q} +\nonumber\\ 
	 J_{ls} \mathbf{(L \otimes
	S)\cdot(l\otimes \sigma)}+ J_{qs} \mathbf{(Q\otimes S)\cdot(q\otimes \sigma)} +J_{p}n.
\end{eqnarray} 
where $ \mathbf{S}, \mathbf{L}, \mathbf{Q}$ are respectively the
impurity spin, orbital, orbital-quadrupole operators and $ \mathbf{s},
\mathbf{l}, \mathbf{q}$ are the corresponding operators for bath
electrons at the position of the impurity, and $n$ is their charge
($J_p$ is the potential scattering parameter).  The five quadrupole
operators $\mathbf{Q}$ are second order orbital tensor operators
defined as $Q^{bc}_{i,j} =
\left(L^{b}_{i,m}L^{c}_{m,j}+L^{c}_{i,m}L^{b}_{m,j} \right)/2 -
\frac{2}{3} \delta_{b,c} \delta_{i,j}$ for $bc=11,12,13,23,33$. 
 The three-orbital conduction band is described
by $H_0$ and assumed flat with its half-bandwidth
$D=1$ taken as the energy unit. The parameters of the Kondo
Hamiltonian can be obtained from the Anderson model by the
Schrieffer-Wolff approximation~\cite{Horvat2016}. In the paper, we
consider values of the Kondo parameters $J_{p}=0.0044, J_{s}=0.025,
J_{l}=0.033, J_{q}=0.035, J_{ls}=0.059, J_{qs}=0.055$ that correspond
to an Anderson-Kanamori model with $U=3.2,J=0.4$ and hybridization
function $\Gamma=0.1$ at the occupancy of $N=2$ electrons. We will
refer to the Kondo model with those parameters as the ``realistic
Kondo model''. In order to reveal the interesting physics we will also
relax the parameters from these values as described in the captions of
the corresponding plots.

We solved the model Eq.~\eqref{Hkk0} with the numerical
renormalization group (NRG)
method~\cite{Bulla2008,ZitkoNRG,Zitko2011}. We took $\Lambda=5$ and
kept up to 3000 states in the diagonalization. We verified that
increasing this number to 4000 and/or varying the value of $\Lambda$
does not affect the results appreciably. We used the $z$-interleaving
with 8 different choices of $z$.

{\it RG equations --} The Hamiltonian has been studied by 
perturbative renormalization group (RG) in
Ref.~\onlinecite{Horvat2016}. One of the main results from that work
is that under the RG flow the difference between quadrupole and
orbital coupling constants becomes unimportant at low energies,
that is $J_{ls}/J_{qs} \rightarrow 1$, $J_{l}/J_{q} \rightarrow 1$,
and the physics becomes that of the problem with higher SU(3) orbital
symmetry~\cite{Aron2015,Horvat2016}.

The RG equations to lowest order for a flat density of states
(additionally, for brevity and clarity,  we take
$J_{ls}/J_{qs}=J_{l}/J_{q}=1$) read
\begin{eqnarray}
\beta_s&=-1/9 (9J_s^2 +8 J_{ls}^2), \\
\beta_l&=-1/8 (12 J_l^2 + 9 J_{ls}^2), \\
\beta_{ls}&=-1/6 (5J_{ls}^2 +12 J_{ls} J_s + 18 J_{ls} J_{l}), \\
\beta_p&=0.
\end{eqnarray}
There are several points worth stressing. (i) The mixed terms $J_{ls}$
drive the spin and orbital coupling constants to $\infty$, hence to a
fully screened Fermi liquid regime. (ii) If the mixed terms are
initially 0, they remain 0 under the RG flow (a conclusion that holds
to all orders, as revealed by the NRG results discussed later). (iii)
For $J_{ls}=0$, the spin and orbital moments in the equations above
decouple. In that limit, the running of orbital coupling constant is
faster, which is associated with a higher SU(3) symmetry. (In contrast
to the RG equations, NRG results show that even for vanishing $J_{ls}$
the spin- and orbital- moments are still coupled, for instance, the
spin-Kondo temperature depends also on $J_l$ and is not simply
exponential in $J_s$ as the equations above suggest.)

\begin{figure}
 \begin{center}
   \includegraphics[width=1.0\columnwidth,keepaspectratio]{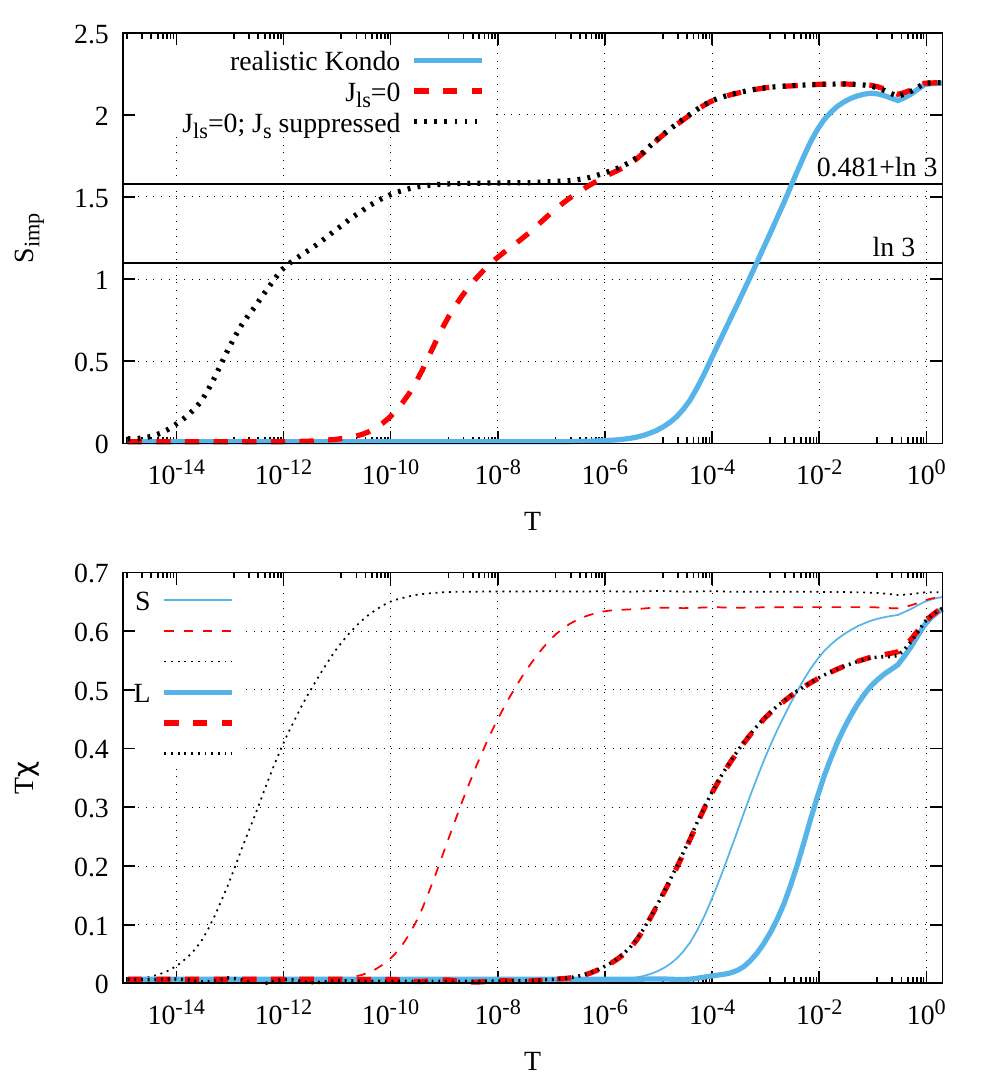}
   \end{center}
\caption{\label{fig1} (top) Impurity contribution to entropy for the
Kondo model corresponding to an Anderson model of a realistic impurity
(full), the case with vanishing mixed couplings $J_{ls}=J_{qs}=0$
(dashed), and the case with additionally suppressed spin-Kondo
coupling from $J_s=0.025$ to $J_{ss}=0.00025$ (dotted). (bottom) The
corresponding orbital (thick) and spin (thin) effective moments. }
\end{figure}

{\it NRG results --}  Fig.~\ref{fig1} shows the impurity
contribution to entropy (top panel) and the spin and orbital
susceptibilities $\chi$ (bottom panel). In the
realistic Kondo model the entropy smoothly diminishes from the value 2$\ln
3$ characteristic of freely fluctuating spin and orbital moments (for
$S=1$, $L=1$), without any pronounced features. 
It is only by looking
separately at the spin and orbital susceptibilities that the two-stage
screening process becomes apparent.  

To reveal the physics more clearly, it is convenient to suppress the
mixed spin-orbital terms. This separates the spin and orbital Kondo
scales further apart so that the intermediate temperature state is
visible also in the impurity contribution to the entropy. A shoulder
appears (dashed curve) that becomes more pronounced and takes the form
of a clear plateau if the spin Kondo constant is further suppressed
(dotted). Now, there is a salient point. Naively, one might expect
that the value of the impurity contribution to entropy at the plateau
to be $\log 3$, corresponding to fluctuating spins. The calculated
entropy at the plateau is, however, larger.

In order to understand the underlying physics, we further
simplify the model. First, because the splitting between quadrupole
and orbital terms is irrelevant in the RG sense, as shown in the
earlier work~\cite{Horvat2016} and as discussed above,  one can
consider a problem with the higher SU(3)
symmetry~\cite{Aron2015,Stadler2015,Stadler2019}. Second, because in
the state of our interest the spins are freely fluctuating, one
can neglect the spin-coupling Kondo constant altogether and set
$J_{s}=0$. The intermediate temperature behavior at the plateau thus
corresponds to a problem described by the interaction Hamiltonian $H_K
= T \cdot t $ where $T,t$ are SU(3) objects. The same conduction band
orbital moment $t$ is, however, realized by two spin channels (spin
plays the role of spectator, as it does not appear in the simplified
$H_K$ explicitly), hence the relevant fixed point is that of the
impurity with SU(3) orbital degree of freedom coupled to two
conduction channels. This problem is overscreened, hence a non-Fermi
liquid.

The general overscreened impurity problem with SU(N) symmetry coupled to $K$
conduction channels was explored in the literature in
detail~\cite{parcollet98}. Equation~(6) of  the cited reference reads
\begin{equation}
  S_\mathrm{imp}=\mathrm{ln} \prod_{n=1}^{Q} \frac{\sin\left[\pi (N+1-n)/(N+K)\right]}{\sin\left[ \pi n /(N+K)\right]},
  \end{equation}
with $N=3,K=2,Q=2$ in the present case, which evaluates to \begin{equation}
  S_\mathrm{imp0}=(1/2) \log\left(\frac{3+\sqrt{5}}{2}\right) \approx 0.481
  \end{equation}
Adding the $\ln(3)$ contribution of the fluctuating spin entropy to
that number one obtains 1.58, which coincides with the value of the
residual entropy at the plateau as shown on Fig.~\ref{fig1}.

Once the fixed point is identified, one knows the scaling of the
response functions. For the fixed point at hand, $\chi(T) \sim
\mathrm{const}$, hence the temperature dependence of local moments
resembles a Pauli response one would expect
for a Fermi liquid. On the other hand the scaling of the response as a function of
frequency is more interesting, namely, one expects~\cite{parcollet98} $\chi_L''(\omega)
\sim\omega^{1/5}$.

\begin{figure}
 \begin{center}
   \includegraphics[width=1.0\columnwidth,keepaspectratio]{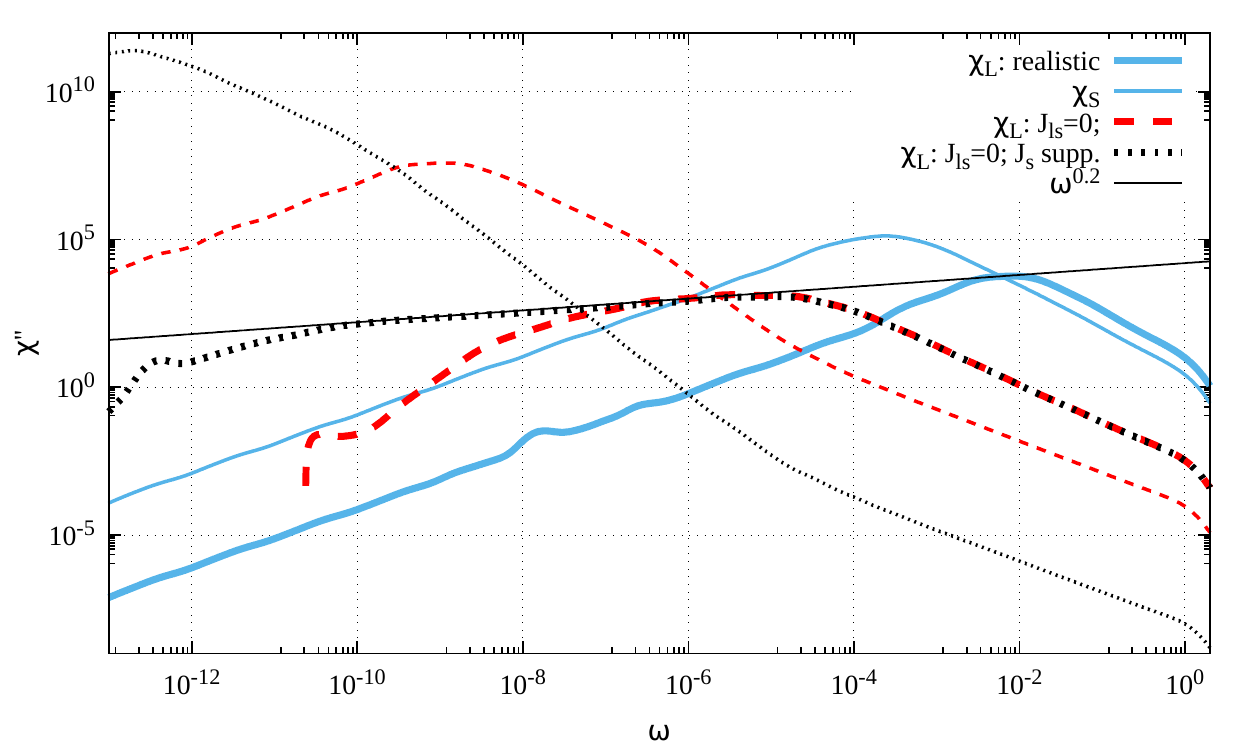}
   \end{center}
 \caption{\label{fig2} Frequency dependence of the imaginary part of spin (thin) and orbital (thick) susceptibilities for parameters as in Fig.~\ref{fig1}. The realistic result was multiplied by a constant. } 
 \end{figure}

Fig.~\ref{fig2} presents the frequency dependence of the imaginary
part of spin and orbital susceptibilities $\chi_{S,L}''(\omega)$,
respectively. The spin susceptibilities at small frequencies
are substantially larger, corresponding to a smaller value of $T_K^S$
($T_K$ can be read directly from the frequency at which the
corresponding susceptibility attains a maximum). In the intermediate
frequency regime $T_K^{S} < \omega < T_K^{L}$ the frequency dependence
of $\chi''_L$ is non-Fermi liquid and follows the $\omega^{1/5}$
dependence, which is particularly clear in curves for the parameters
with suppressed mixed and spin-Kondo couplings. This demonstrates that
the over-screened SU(3) fixed point controls the electron response in
this intermediate energy regime.

\begin{figure}
 \begin{center}
   \includegraphics[width=1.0\columnwidth,keepaspectratio]{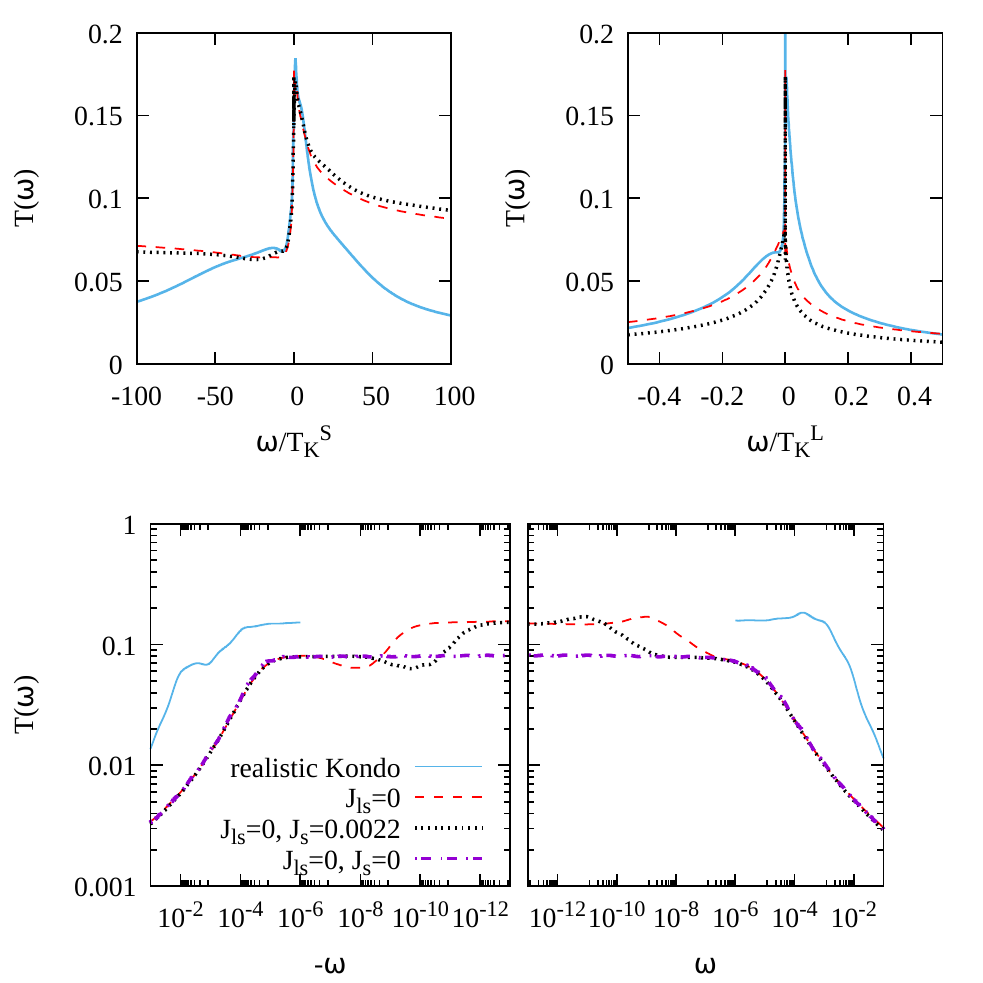}
   \end{center}
 \caption{\label{fig4} Frequency dependence of the spectral function $T(\omega)$. 
}
 \end{figure}

{\it Spectral functions --} Whereas the frequency dependence of the
orbital susceptibility reveals the non-Fermi liquid fixed point in
the most direct way, it is not a quantity that is easily
measured experimentally. Hence we also calculated the impurity
spectral function (that is, the $T$-matrix). We plot the results in
Fig.~\ref{fig4} for the Kondo model with realistic parameters obtained
from the Schrieffer-Wolff transformation and for the case with
suppressed mixed terms, $J_{ls}=J_{qs}=0$. The result for realistic
parameters reveals asymmetric shape of the quasiparticle peak with a
characteristic shoulder (or side-peak) at negative frequency (for the
case corresponding to occupancy of two electrons). The shape of
the quasiparticle peak was discussed in 
Ref.~\onlinecite{Stadler2019} as a narrow needle associated with the
screening of spin, on a broader hump characteristic of the orbital
screening. In earlier work on the Anderson-Kanamori and
Hubbard-Kanamori model~\cite{jm16} it was shown that the frequency of
the side-peak scales with the coherence scale, that is, for increasing
Hubbard repulsion and/or Hund's coupling the feature moves to lower
frequencies, which implies that it is not associated with atomic
satellites. This inner structure of the quasiparticle peak was (to our
knowledge) first discussed in Ref.~\cite{wadati14}, and is seen in the
Anderson-Kanamori model~\cite{Stadler2015}.  

The present results demonstrate that the side feature is present
already in the Kanamori-Kondo model. To explore this in more detail we
plot the spectral function on a logarithmic scale (lower panel)
for a set of $J_s$, including the case of $J_s=0$ that has a
non-Fermi-liquid ground state. One sees that the results for
finite $J_s$  follow the dependence of the $J_s=0$  case until
approaching the spin-Kondo screening scale. At negative frequencies
below the $T_K^{S}$ the spectral function is first suppressed and then
increases as the Fermi liquid coherent state is established.

On this point we also notice that in the case of realistic LDA+DMFT
calculation on Sr$_2$RuO$_4$ the  side-peak is observed at positive frequencies and
that its existence was invoked to explain the measured optical
conductivities\cite{stricker2014}. Hence this structure of the
quasiparticle peak, that is not present for vanishing Hund's rule
coupling, can be considered as a spectral fingerprint of the Hund's metals.

\begin{figure}
 \begin{center}
   \includegraphics[width=1.0\columnwidth,keepaspectratio]{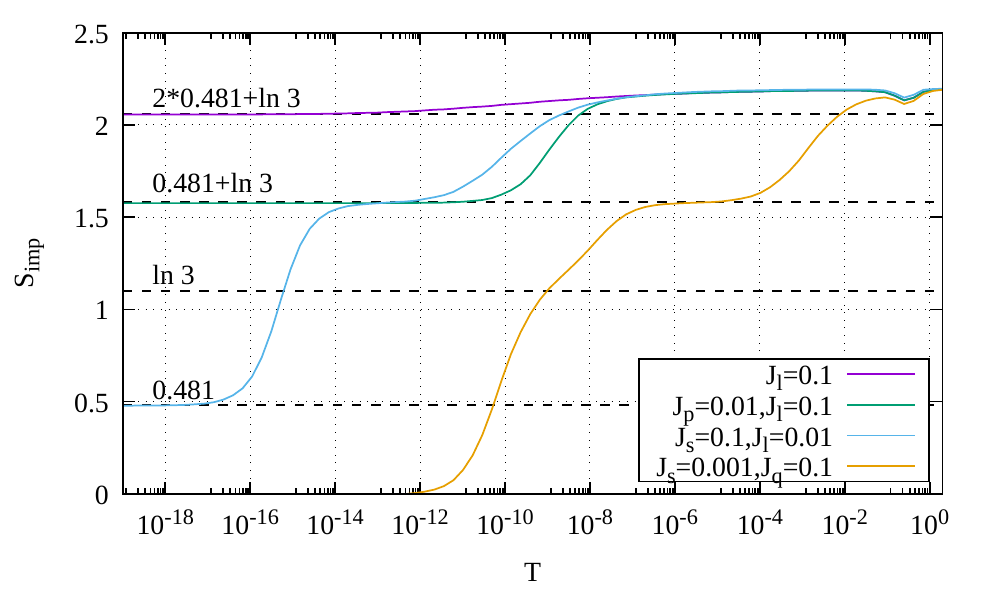}
   \end{center}
\caption{\label{fig3} Temperature dependence of the impurity
contribution to entropy. The non-vanishing parameters for respective
curves are stated in the legend, the parameters that are not stated
explicitly are set to 0.} 
\end{figure}

{\it Other non-Fermi liquid regimes --} Relaxing  the parameters
even further one can access additional non-Fermi liquid regimes. The
impurity contribution to entropy for  some of the interesting
cases is shown on Fig.~\ref{fig3} and the corresponding fixed points are listed in Table~\ref{table1}.  One of the interesting fixed
points discussed first in Ref.~\cite{leo04thesis,leo05} is that of the
case where the quadrupole terms in the Kondo Hamiltonian are
suppressed $J_q=J_{qs}=J_{ls}=0$. In that case,  the
problem has an additional particle-hole symmetry which reflects the
fact that the same spin and orbital moments can be obtained either by
2 or 4 electrons in the impurity (and the matching configuration of
bath electrons). This particle-hole symmetry is broken by explicit
potential scattering $J_p n$ and also by quadrupole terms $J_q
\mathbf{Q}\cdot \mathbf{q}$. 

\begin{table}[h]
\caption{\label{table1} NFL fixed points relevant to Fig.~\ref{fig3};  $S_\mathrm{imp0}=(1/2) \log\left(\frac{3+\sqrt{5}}{2}\right) \approx 0.481.$  }
    \begin{tabular}{c|c|c}
     when   & residual entropy & case  \\\hline
    $J_{l(q)s}=0$, $J_s=0$ & $\ln 3$ + $S_\mathrm{imp0}$ & two-ch. SU(3)+free spin \\
    $J_{q}=J_{l(q)s}=J_p=0$ &   $S_\mathrm{imp0}$ & three-ch. SU(2) \\
only $J_l>0$  & 2$S_\mathrm{imp0} +\ln 3$ & combination of two above 

    \end{tabular}

\end{table}

When those terms are suppressed, the low-energy physics (once the spin
is screened) is that of a hypercharge $Y=1/2$ coupled to three
conduction bands given by orbital degrees of freedom and hence to a
three-channel over-screened $Y=1/2$ problem~\cite{leo04thesis}. For that problem,
the residual entropy evaluates to $S_\mathrm{imp1}=0.481$ (which is,
incidentally, the same as earlier discussed $S_{\mathrm{imp0}}$), and
is the zero-temperature value residual entropy for the case of
$J_s=0.1,J_l=0.01$ of Fig.~\ref{fig3}. In that case, the spin is fully
screened, but the hyper-charge fluctuations lead to the physics just
discussed.  If the quadrupole terms are retained (case
$J_s=0.1,J_q=0.01$), the particle-hole symmetry is broken and this
physics is not realized.  If one suppresses the spin Kondo coupling
constant (case where only $J_l=0.1$ is non-vanishing), the spin is
freely fluctuating (the contribution to entropy $\mathrm{ln}3$), and
one has a simultaneous occurrence of two-channel over-screened SU(3)
problem in the orbital sector and the three-channel over-screened
SU(2) problem in the hyper-charge sector, hence one reaches $2\times
0.481 +\mathrm{ln} 3$ residual entropy. Adding the potential
scattering (case $J_p=0.01,J_l=0.1$) suppresses the particle-hole
symmetry, and the residual entropy is that of the over-screened
problem in the orbital sector only.

Studying the frequency dependencies of the corresponding correlation
functions for the cases just discussed is an interesting subject for
the future work.

{\it Summary -- } We have revealed a non-Fermi-liquid fixed point that
corresponds to the idealized incoherent state of three-orbital Hund's
metals within the DMFT description. At energies controlled by this
fixed point the orbital susceptibility has an unusual $\omega^{1/5}$
dependence and the spectral function is roughly constant which
manifests as a side peak at negative (positive) frequencies for
$N_d=2$ ($N_d=4$, related to the optical observations on Sr$_2$RuO$_4$
\cite{stricker2014}). It may be that because within the Anderson model
the mixed terms are necessarily larger $J_{ls}/J_{l}>1$ the revealed
fixed point is not approached closely, but also in that case the
presented results provide a useful precise reference point for further
discussion.

 It would be interesting to develop experimental
techniques to investigate the frequency dependent orbital
susceptibility, for instance in
ruthenates, especially on approaching the half-filled configuration
where one can expect a larger separations between the spin- and the
orbital screening scales~\cite{Horvat2016}.


\begin{acknowledgments}
A.~H., R.~\v{Z}., and J.~M. are supported by Slovenian Research Agency
(ARRS) under Program P1-0044 and Project J1-7259. We warmly thank Antoine Georges for the discussions in particular the ones related to Ref.~\cite{parcollet98}.
\end{acknowledgments}

\bibliography{kondo.bib}

\end{document}